\documentclass[fleqn,usenatbib]{mnras}

\usepackage{newtxtext,newtxmath}

\usepackage[T1]{fontenc}

\DeclareRobustCommand{\VAN}[3]{#2}
\let\VANthebibliography\thebibliography
\def\thebibliography{\DeclareRobustCommand{\VAN}[3]{##3}\VANthebibliography}

\usepackage{graphicx}	
\usepackage{amsmath}	
\usepackage{amssymb}	

\title[Star-Forming S0 Galaxies in SDSS-IV MaNGA Survey]{Star-Forming S0 Galaxies in SDSS-IV MaNGA Survey}

\author[X. Ke et al.]{
Ke Xu,$^{1,2}$
Qiusheng Gu,$^{1,2}$\thanks{E-mail: qsgu@nju.edu.cn}
Shiying Lu,$^{1,2}$
Xue Ge,$^{3}$
Mengyuan Xiao$^{1,2}$
and Emanuele Contini$^{1,2}$
\\
$^{1}$School of Astronomy and Space Science, Nanjing University, Nanjing 210093, P. R. China\\
$^{2}$Key Laboratory of Modern Astronomy and Astrophysics (Nanjing University), Ministry of Education, Nanjing 210093, China\\
$^{3}$School of Physics and Electronic Engineering, Jiangsu Second Normal University, Nanjing, Jiangsu 211200, China
}

\date{Accepted XXX. Received YYY; in original form ZZZ}

\pubyear{2021}

\begin{document}
\label{firstpage}
\pagerange{\pageref{firstpage}--\pageref{lastpage}}
\maketitle

\begin{abstract}
To investigate star-forming activities in early-type galaxies, we select a sample of 52 star-forming S0 galaxies (SFS0s) from the SDSS-IV MaNGA survey. We find that SFS0s have smaller stellar mass 
compared to normal S0s in MaNGA. After matching the stellar mass to select the control sample, we find that the mean S\'{e}rsic index of SFS0s' bulges (1.76$\pm$0.21) 
is significantly smaller than that of the control sample 
(2.57$\pm$0.20), suggesting the existence of a pseudo bulge in SFS0s. After introducing the environmental information, SFS0s show smaller spin parameters in the field than in groups, while the control sample has no obvious difference in different environments, which may suggest different dynamical processes in SFS0s. Furthermore, with derived N/O and O/H abundance ratios, SFS0s in the field show nitrogen enrichment, providing evidence for the accretion of metal-poor gas in the field environment. 
To study the star formation relation, we show that the slope of the spatially resolved star formation main sequence is nearly 1.0 with MaNGA IFU data, confirming the self-regulation of star formation activities at the kpc scales.
\end{abstract}




\section{Introduction}\label{sec:intro}

The galaxy morphology encodes the clues of galaxy formation and evolution. The famous `tunning fork' \citep{1936rene.book.....H}
separates spirals and ellipticals on the two sides of lenticular galaxies (S0s). S0 galaxies contain the disc structure as spirals, except for the more prominent bulges and absence of the spiral patterns, 
which are the significant bridge connecting early-type galaxies (ETGs) and late-type galaxies (LTGs). S0s and ellipticals are usually classified as ETGs. Because ETGs are redder than LTGs, the gas evacuation 
mechanism is required to suppress star formation and 
preserve the old stellar population \citep[e.g.,][]{2001AJ....122.1861S}, resulting in no recent star formation and no cold 
gas in ETGs.\\
\indent However, recent observations indicate that some ETGs contain atomic gas, and even dust and/or molecular gas \citep[e.g.,][]{2006MNRAS.371..157M,1977A&A....60L..23B,2001AJ....121..808C,2007MNRAS.377.1795C}. The related star formation activities are revealed, e.g., from far-ultraviolet and infrared radiation \citep{2014MNRAS.444.3427D}, or from the emission-line diagnostic analysis \citep{2016ApJ...831...63X}. 
Yet, some ETGs with higher gas surface density than normal spirals fall systematically below  Kennicutt-Schmidt law \citep{1959ApJ...129..243S,1998ApJ...498..541K} and Elmegreen-Silk relation 
\citep{1997RMxAC...6..165E,1997ApJ...481..703S}, showing lower star formation efficiency than LTGs \citep{2014MNRAS.444.3427D}. Because of the above facts, \citet{2014MNRAS.444.3427D} used the 
local volumetric star formation relation 
to unify the star formation of spirals, starbursts and ETGs, and proposed the dynamical origin of the star formation suppression in ETGs.\\
\indent On the other hand, different environments may have different imprints during the formation and evolution of S0s. \citet{2020MNRAS.498.2372D} utilized the SAMI survey and proposed two formation pathways 
for S0s. One is through minor mergers in the field, and the other is from the faded spirals in the denser environment. The different star formation history could lead to 
different features of S0s. While the faded spirals 
preserve more rotation-supported discs \citep[e.g., ][]{2020MNRAS.498.2372D}, 
according to numerical simulations, minor mergers, frequently happening in the local universe \citep[e.g.,][]{2008MNRAS.391.1806V}, have been shown to have the effect of reducing the specific angular momentum, thickening the disc, and contributing to the random motion \citep{2021MNRAS.502.3085G}. The specific angular momentum is an indispensable quantity influencing the atomic gas component  in a disc galaxy \citep[e.g.,][]{2016ApJ...824L..26O}, preventing the gas from infalling and suppressing star formation. As for galaxies with lower specific angular momentum, the corresponding low Toomre parameter Q \citep{1964ApJ...139.1217T} would lead to disc instability and consume gas more efficiently within a time scale $\sim$ 2-3 Gyr \citep{2018MNRAS.480L..23R,2019MNRAS.483.2398M,2020MNRAS.491.4843R,2020MNRAS.499.5656R}, which is significantly shorter than the cosmic time, remaining the galaxy deficient in atomic gas and lack of the raw material for further star formation. But for S0s formed via minor mergers recently, they are still dwelling in star formation conditions, and would provide clues about physics in star formation activities triggered by mergers or accretions.\\
\indent Today, to study S0s with undergoing star formation activities is of interests, which will not only help to understand the process of the quenching or rejuvenation in ETGs, but also help to figure out the nature of star formation law in the case of ETGs. Nowadays, with available Integral Field Unit (IFU) data, such as MaNGA, it is possible to analyze the spatially resolved dynamics and chemical components, providing valuable information on the formation and evolution of S0s.\\
\indent In this paper, we have two main purposes. The first is to obtain the basic information of the star-forming S0s (hereafter SFS0s), such as stellar mass, S\'{e}rsic index of bulges, and star formation main sequence (SFMS). The second is to study the relation between star formation activities and environments.\\
\indent This paper is organized as follows: In Section \ref{sec:data}, we introduce sample selection, basic physical parameters and preliminary analysis. In Section \ref{sec:results}, we describe the dynamics and chemical enrichment of S0s in different environments. We also discuss the origin of star-forming S0s in Section \ref{discussion}, and conclude in Section \ref{sec:conclusion}. In this paper, we carry out our work using Salpeter Initial Mass Function \citep{1955ApJ...121..161S} in a flat $\rm\Lambda CDM$ cosmology, with parameters: $\rm\Omega_M=0.3$, $\rm\Omega_{\Lambda}=0.7$ and $\rm H_0=70\,km\cdot s^{-1}\cdot Mpc^{-1}$.

\section{Data} \label{sec:data}

In this work, we utilize the data products from MaNGA \citep{2015ApJ...798....7B,2016AJ....152..197Y} Data Release 16 \citep{2020ApJS..249....3A}. We use the `MAPS-VOR10-GAU-MILESHC' files from the Data Analysis Pipeline \citep[DAP,][]{2019AJ....158..231W,2019AJ....158..160B} for emission-line information and stellar kinematics. The DAP pipeline simultaneously fits the emission line and stellar continuum, minimizing the stellar features on the recovered emission-line fluxes.  We use PIPE3D value-added catalogue  \citep{2016RMxAA..52...21S,2016RMxAA..52..171S} for stellar mass surface density, redshift, axis ratio, and ionization information in the central 2.5-arcsecond aperture. The total stellar mass and star formation rate are taken from the MPA-JHU catalogue \citep{2003MNRAS.341...33K,2004MNRAS.351.1151B}.

\subsection{Sample Selection} \label{sec:sampleselection}

\begin{figure}
\centering
\includegraphics[width=0.5\textwidth]{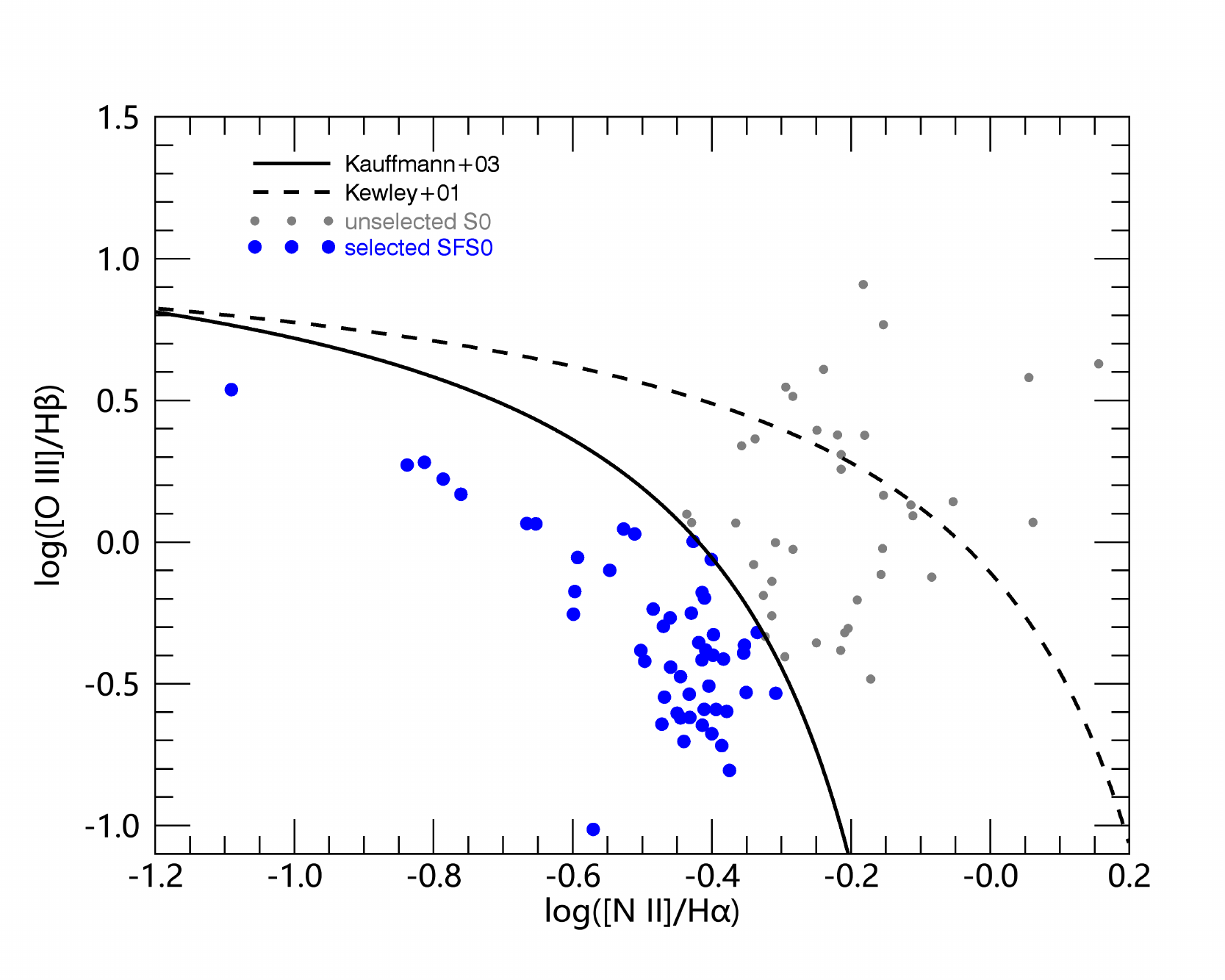}
\caption{The BPT diagram for SFS0s. The blue (grey) dots are the selected SFS0s (unselected S0s). The solid and dashed demarcation lines are taken from \citet{2001ApJ...556..121K} and \citet{2003MNRAS.346.1055K}.\label{fig:1}}
\vspace{0.2cm}
\end{figure}

\begin{figure}
\centering
\includegraphics[width=0.5\textwidth]{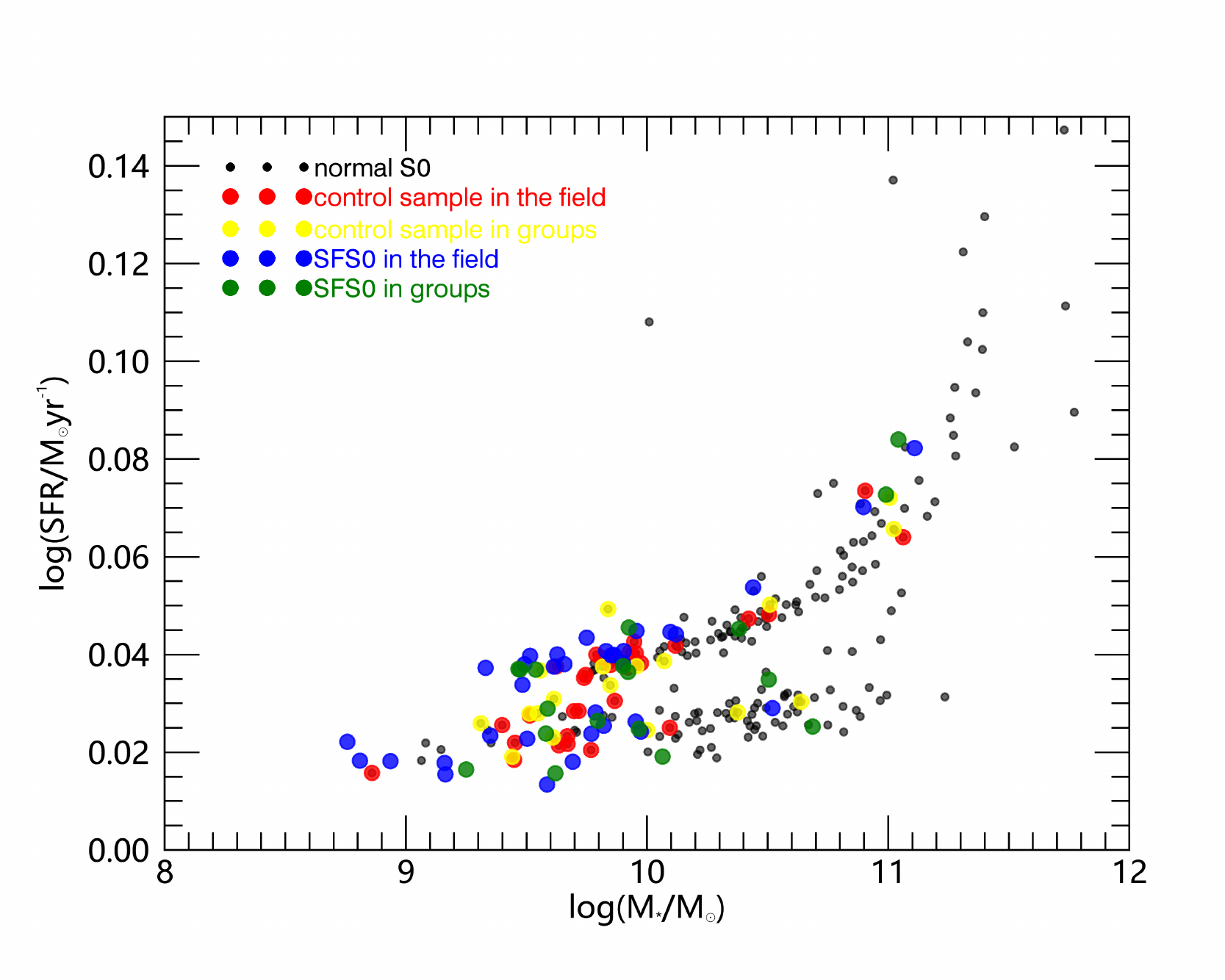}
\caption{The stellar mass vs redshift distributions. The dots are: normal S0s (black), the control sample in the field (red), the control sample in groups (yellow), SFS0s in the field (blue), SFS0s in groups (green). 
\label{fig:2}}
\vspace{0.2cm}
\end{figure}

\citet{2019MNRAS.483.2057F} described the Sloan Digital Sky Survey IV (SDSS-IV) MaNGA Pymorph Photometric Value-Added Catalogues (MPP-VAC) and MaNGA Deep Learning Morphology Value-Added Catalogues (MDLM-VAC). They included the photometric parameters and morphological classification, which have been eye-balled, providing valuable morphological information for later analysis.\\
\indent In this work, we first select galaxies in MDLM-VAC with TType $<$ 0, ensuring them to be early-type galaxies. Secondly, these galaxies are required to be S0 candidates, with P\_S0 $>$ 0.7. Thirdly, the ellipticity is required less than 0.7 to avoid the edge-on case. Fourthly, to decide whether the galaxy has ongoing star formation activities, we perform the traditional BPT diagnostic \citep{1981PASP...93....5B} and restrict the S/N $>$ 5 for all emission lines in the central 2.5-arcsecond aperture. The demarcation lines are taken from \citet{2001ApJ...556..121K} and \citet{2003MNRAS.346.1055K} to select star-forming S0s. We further require the H$\rm \alpha$ equivalent width in the central 2.5-arcsecond aperture to be larger than 6\AA\;\citep[e.g., ][]{2020ARA&A..58...99S}. The corresponding diagnostic is shown in Fig.~\ref{fig:1}. After excluding AGNs and composite types, we obtain a sample of 52 SFS0s. Finally, a sample of 216 normal S0s are also constructed by adopting the criterion as in \citet{2020MNRAS.499..230B}, 1.1 dex lower than the SFMS given by \citet{2015ApJ...801L..29R}. The stellar mass vs redshift distribution of normal S0s and SFS0s are plotted in Fig.~\ref{fig:2}. The two sequences in Fig.~\ref{fig:2} are Primary and Secondary samples in the MaNGA survey \citep[][]{2015ApJ...798....7B,2016AJ....152..197Y}, which reach 1.5 and 2.5 $\rm R_e$ for more than 80$\rm\%$ of the galaxies individually. The main conclusions are compatible in the two sequences, and 
we will use both samples in this work.\\
\vspace{0.6cm}
\vspace{0.6cm}

\subsection{Environmental information} \label{sec:environment}
As discussed in \citet{2020MNRAS.498.2372D}, the environment plays an essential role in the formation and evolution of S0 galaxies, leading to distinct origins in the field and denser groups. 
\citet{2007ApJ...671..153Y} presented the group catalogue for SDSS galaxies. By cross-matching our galaxies with their catalogue, we separate our sample as \citet{2020MNRAS.498.2372D}: 
isolated S0s which are not associated with any group, grouped S0s 
which are in groups with halo mass larger than 10$^{11}$M$_{\odot}$. We finally get 52 SFS0s (34 in the field and 18 
in groups) and 208 normal S0s (108 in the field and 100 
in groups).

\subsection{Control sample}\label{sec:cs}
\begin{figure}
\centering
\includegraphics[width=0.5\textwidth]{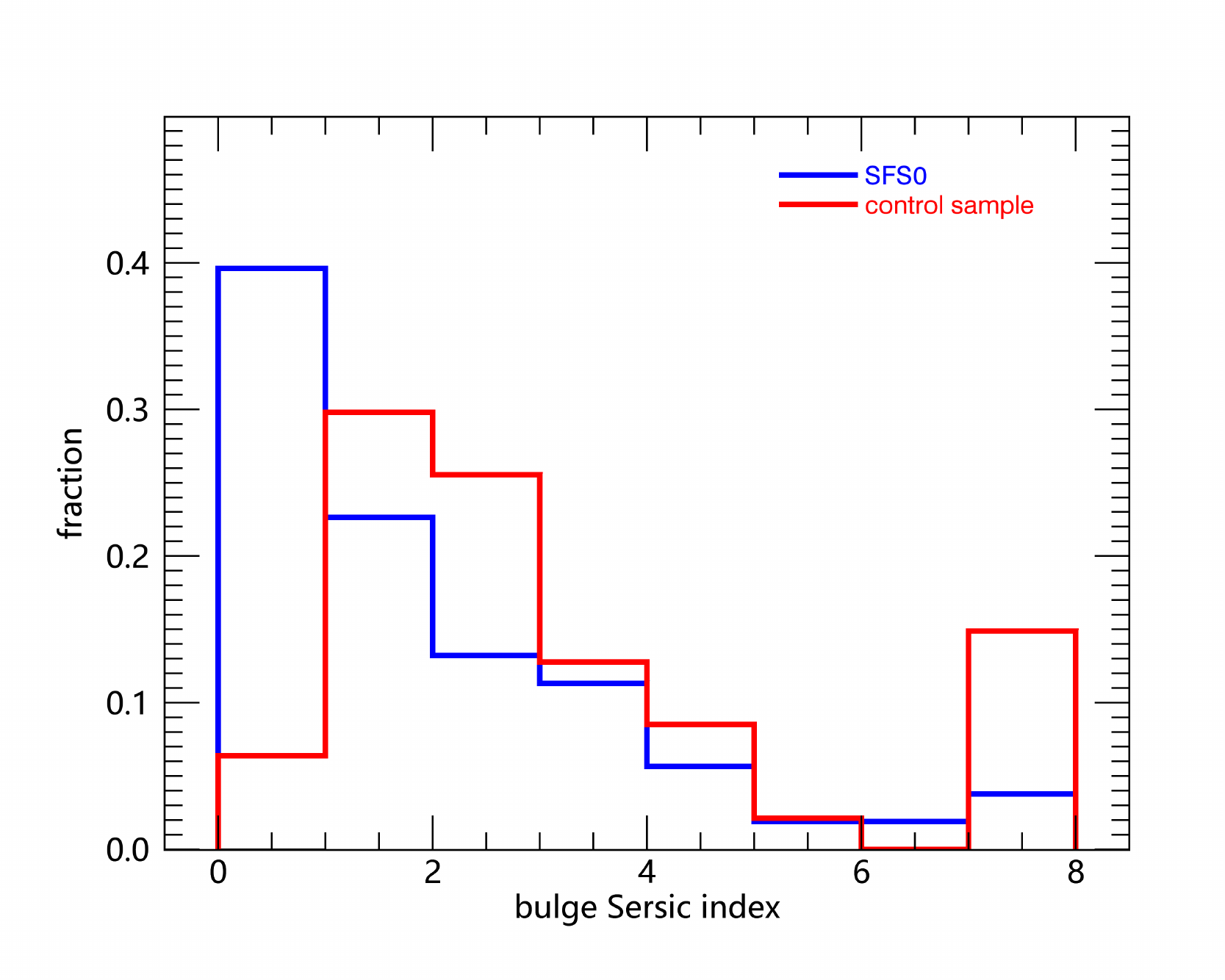}
\caption{The distributions of the S\'{e}rsic index of bulges. Blue lines represent SFS0s, and red lines represent the control sample. The excess at n=8 is due to the upper limit in the morphological fitting.\label{fig:4}}
\vspace{0.2cm}
\end{figure}

Fig.~\ref{fig:2} shows that normal S0s present higher stellar mass, which may be related to the mass quenching discussed in the previous works \citep[][]{2016ApJ...831...63X,2020ApJ...902...75L}. For comparison, a control sample with similar stellar mass distribution as SFS0 is required. For each SFS0, we match a control galaxy from the normal ones with $\rm\Delta log(M_*)<0.15$ and $\rm\Delta z<0.02$. In this way, a sample of 48 normal S0s are selected, while 4 SFS0s fail to match a counterpart due to the lack of quiescent S0s below $\rm log(M_*/M_{\odot})<9.5$. Removing these 4 SFS0s would not affect our results, and we preserve 52 SFS0 sample and 48 normal S0s as the control sample. Fig.~\ref{fig:2} shows their stellar mass vs redshift distributions.\\
\indent To have a better knowledge of the two samples, we compare their morphologies. We plot the bulge S\'{e}rsic index from MPP-VAC of SFS0s and the control sample in Fig.~\ref{fig:4}. Note that the excess at S\'{e}rsic index n = 8 is due to the upper limit in the fitting procedure. Regardless of this, SFS0s and the control sample have mean values of 1.76$\pm$0.21 and 2.57$\pm$0.20, respectively, indicating that SFS0s contain more pseudo bulges than the control sample, consistent with \citet{2016ApJ...831...63X}. 
\citet{2013MNRAS.432.1862C} found the inverse relation between bulge fraction and molecular gas content. A classic bulge would lead to less cold gas at the centre. In contrast, the shallower potential well for pseudo bulges introduces the weaker dynamical suppression as \citet{2014MNRAS.444.3427D} suggested.

\vspace{0.6cm}

\subsection{Spin parameter} \label{sec:dynamics}
From MaNGA DAP data, we obtain the spatially resolved radial stellar velocity and stellar velocity dispersions. 
The velocity dispersion data account for the effect of beam smearing and instrumental dispersion following \citet{2020ApJ...893...26S} and \citet{2019AJ....158..231W}. The effective radius and ellipticity are from PIPE3D for recovering the rotational velocity.
The position angle is determined along the largest velocity gradient. We only analyze the 
spaxels within 45 degrees from the major axis of the disc, avoiding the divergence near the minor axis \citep[similar method as in][]{2020MNRAS.498.2372D}. 
We estimate the inclination from the ellipticity:
\begin{equation}
{\rm cos}\,i=\sqrt{\frac{(1-\epsilon)^2-q_0^2}{1-q_0^2}},
\end{equation}
where $i$ is the inclination, $\epsilon$ is the ellipticity, and $q_0$ is the intrinsic axis ratio of edge-on galaxies, which is assumed to be 0.2 \citep{2000ApJ...533..744T}. We reconstruct the spatially resolved rotational velocity as in \citet{2020MNRAS.498.2372D}: 
\begin{equation}
v_t=\frac{v_r}{{\rm sin}\,i\,{\rm cos}\,\theta},
\end{equation}
where $\theta$ is the azimuthal angle from the major axis, $v_t$ and $v_r$ are the rotational and radial velocities in each spaxel, respectively.\\
\indent Following \citet{2007MNRAS.379..418C}, the spin parameter within 1.5 $\rm R_e$ is defined as:
\begin{equation}
\lambda_R=\frac{\sum_{n=1}^NF_nR_n\vert v_t \vert}{\sum_{n=1}^NF_nR_n\sqrt{v_t^2+\sigma_n^2}},
\end{equation}
where n denotes the nth spaxel, $F_n$ is the flux, $R_n$ is the distance from the centre, and $\sigma_n $ is velocity dispersion. The rotational velocity ($v_t$) is used to minimize the influence of inclination.

\subsection{Star-formation rate} \label{sec:sfr}
From the luminosity of the H$\alpha$ emission line, we calculate the spatially resolved star-formation rate (SFR) in each spaxel of SFS0s. Only the spaxels with H$\rm\alpha$ equivalent length $>$ 6\;\AA\;are included \citep{2020ARA&A..58...99S}.
First, we adopt the extinction correction following the method by \citet{2020MNRAS.492...96B}:
\begin{equation}
A(X)=-2.5\times {\rm log}(\frac{f_{\rm H\alpha}/f_{\rm H\beta}}{2.86})\times \frac{K_X}{K_{\rm H\alpha}-K_{\rm H\beta}},\quad {\rm and}
\end{equation}
\begin{equation}
K_X\equiv A(X)/A(V),
\end{equation}
where $A(X)$ is the dust reddening at wavelength X. Assuming the intrinsic Balmer decrement of $\rm H\alpha/H\beta=2.86$, we derive the corrected emission line flux ($f_{X,{\rm corr}}$) by:
\begin{equation}
f_{X,{\rm corr}}=f_{X,{\rm obs}}\times 10^{A(X)/2.5},
\end{equation}
where $f_{X,{\rm obs}}$ is the observed flux. We adopt $R_V=3.1$ and the extinction curve of \citet{1989ApJ...345..245C}. All emission lines are extinction-corrected. After converting the flux into luminosity, the SFR in each spaxel is finally calculated from the relation given by \citet{1998ApJ...498..541K}:
\begin{equation}
{\rm SFR}_{\rm H\alpha}[{\rm M_{\odot}/yr}]=7.9\times10^{-42}L_{\rm H\alpha}[{\rm erg\cdot s^{-1}}].
\end{equation}

\subsection{Gas inflow indicator}\label{sec:gasinflow}
\citet{2021ApJ...908..183L} used N/O abundance ratio to 
study the accretion of metal-poor gas in star-forming galaxies. The basic idea is that oxygen is the primary element depending on stellar nuclear activities, while nitrogen can be the primary and secondary element. In a low-metallicity star, the helium-burning provides materials for nitrogen and oxygen production, leading to a constant N/O ratio, while a high-metallicity star has carbon and oxygen from ISM, resulting in N/O dependent on O/H ratio \citep[detailed in][]{2021ApJ...908..183L}. The different roles of nitrogen determine its relationship between N/O and O/H \citep[e.g.,][]{2013ApJ...765..140A}. When galaxies accrete the metal-poor gas, they would locate above the relation between N/O and O/H, resulting in N/O abundance excess. 
During the transport of gas into galaxies, star formation is sustained.\\ 
\indent Does it happen in SFS0s? We follow the same procedure to probe the corresponding properties and restrict the spaxel with H$\rm\alpha$ equivalent length $>$ 6\;\AA\; and emission lines with S/N $>$ 5 to calculate the metallicity. We adopt the relation in \citet{2020ApJ...891...19L} to calculate N/O:
\begin{equation}
\begin{aligned}
&{\rm log(N/O)}=0.73\times {\rm N2O2}-0.58,\quad{\rm and}
\\
&\rm N2O2=[NII]\lambda 6584/[OII]\lambda\lambda 3727,3729.
\end{aligned}
\end{equation}
Instead of using the common O3N2 method in consideration of degeneracy, the O/H is obtained from the RS32 method in \citet{2020MNRAS.491..944C}, which is only mildly dependent on ionization parameter:
\begin{equation}
\begin{aligned}
&{\rm RS_{32}}=\sum\nolimits_{\rm N}c_n[12+{\rm log}({\rm O/H})-8.69]^n,\quad {\rm and}
\\
&\rm RS_{32}=[OIII]\lambda 5007/H\beta + [SII]\lambda\lambda 6717, 31/H\alpha,
\end{aligned}
\end{equation}
where $c_n$ is from Table 2 in \citet{2020MNRAS.491..944C}.

\section{Results} \label{sec:results}

\subsection{Star formation main sequence}\label{sec:rSFMS}
\begin{figure}
\centering
\includegraphics[width=0.5\textwidth]{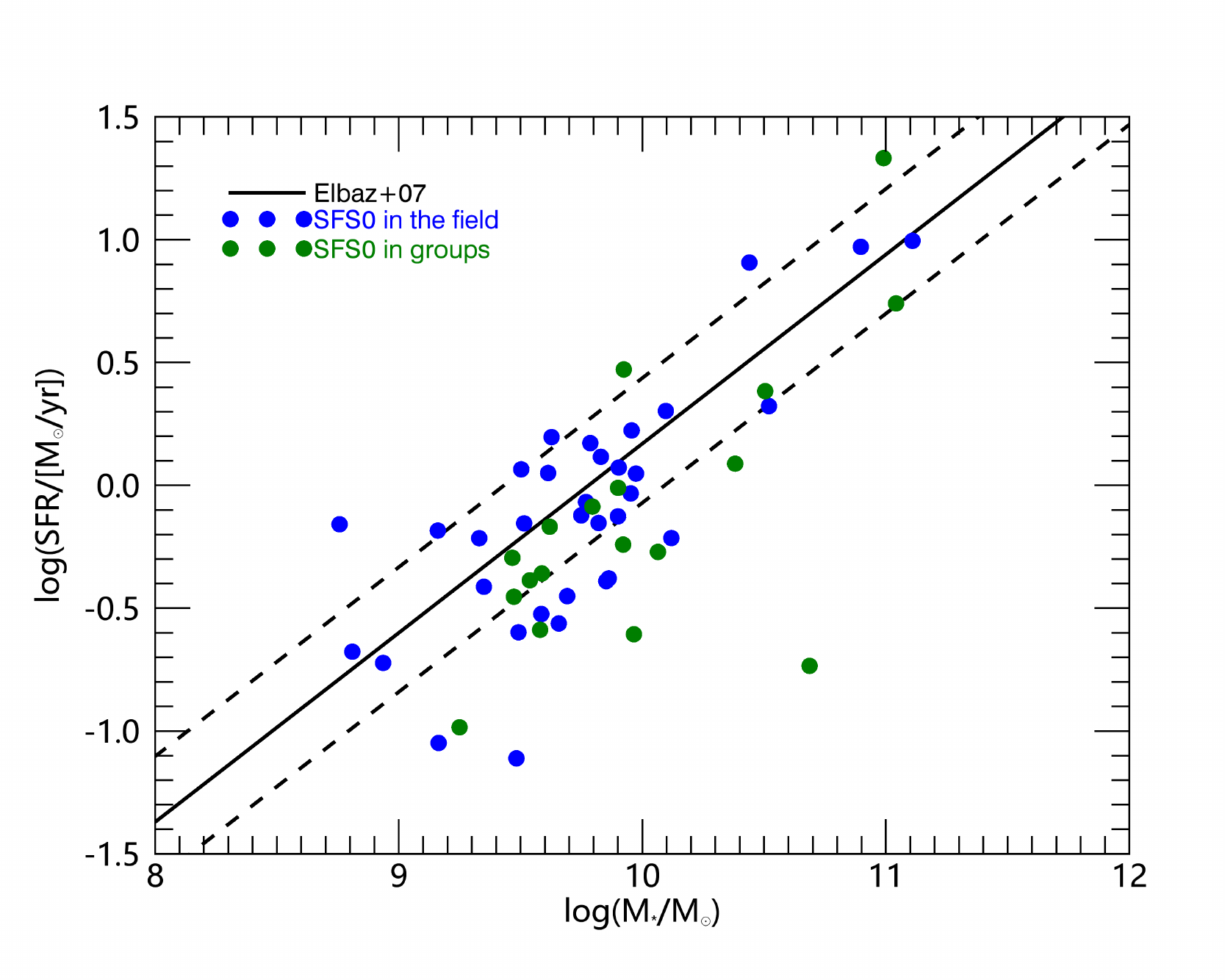}
\caption{The star formation main sequence of SFS0s. The blue points are isolated SFS0s, and the green points are SFS0s groups. The solid line is from \citet{2007A&A...468...33E}, and the two dashed lines correspond to $\pm$1$\sigma$ thresholds.\label{fig:5}}

\end{figure}

\begin{figure}
\centering
\includegraphics[width=0.5\textwidth]{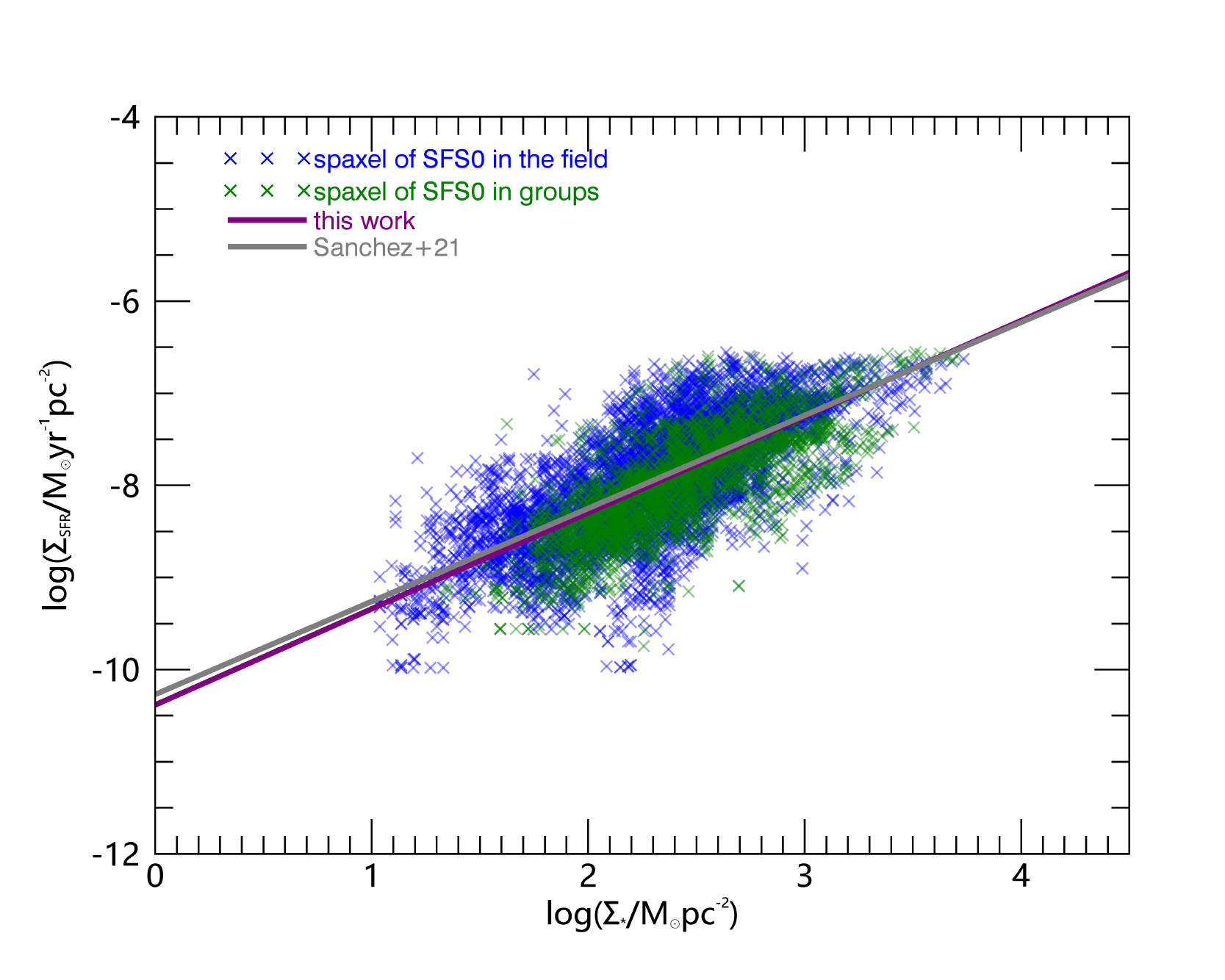}
\caption{The spatially resolved SFMS for our SFS0s. Blue crosses: star-forming spaxels from SFS0s in the field. Green crosses: star-forming spaxels from SFS0s in groups. The solid purple line is the best-fitting of all the spaxels in this work, while the solid grey line is the fitting of the CALIFA dataset from \citet{2021MNRAS.503.1615S}.\label{fig:6}}
\vspace{0.2cm}
\end{figure}

In Fig.~\ref{fig:5}, we plot the SFMS only for SFS0s, and due to the absence of emission lines, galaxies of the control sample are not shown. The total stellar mass and star formation rate are derived from the MPA-JHU catalogue. The solid line in Fig.~\ref{fig:5} is from \citet{2007A&A...468...33E}, and describes the SFMS of SDSS $z\sim 0$ galaxies with the same data from the MPA-JUH catalogue. 10/34 of SFS0s in the field and 7/18 of SFS0s in groups are below 1$\sigma$ threshold of the relation. While the SFS0s are selected according to the central 2.5-arcsecond aperture, most locate well on the SFMS, suggesting a relatively steady star formation rate as star-forming disc galaxies. About 1/3 SFS0s lower than the SFMS might be transforming between blue and red sequences. Because some galaxies would scatter below the 1-$\rm\sigma$ limit due to statistical fluctuations, 1/3 is probably overestimated.\\
\indent Considering smaller physical scales, \citet{2021MNRAS.503.1615S} also presented the resolved SFMS (rSFMS) relation based on the different data sets in the local universe (e.g., for the CALIFA survey, the slope and intercept are 1.01$\pm$0.15 and -10.27$\pm$0.22, respectively). Our work gives the rSFMS relation from the SFS0 spaxels with H$\rm\alpha$ equivalent width $>$ 6\;\AA\; (detailed in Sec.~\ref{sec:sfr}). The stellar mass in each spaxel is from the PIPE3D pipeline. The spaxel area is derived from redshift and spaxel angular size accounting for galaxy inclination. The results are presented in Fig.~\ref{fig:6}. Because the fitting of spaxels from the SFS0s in the field and groups are similar, we show the fitting result combining all the spaxels, where the derived slope and intercept are 1.04$\pm$0.01 and -10.38$\pm$0.03, respectively, nearly the same as \citet{2021MNRAS.503.1615S}. The existence of such rSFMS relation implies that star formation in SFS0s is self-regulated by the same physical process at kpc scales as normal star-forming galaxies \citep{2021MNRAS.503.1615S}. Given the nature of S0 formation via faded spirals in groups or minor mergers and accretions in the field, this further confirms that the galaxy interactions with themselves or among the environments do not affect star formation globally wide, but through regulation at resolved scales \citep[][]{2020ARA&A..58...99S,2021MNRAS.503.1615S}. 

\subsection{Spin parameter distribution} \label{sec:spin}
\begin{figure}
\centering
\includegraphics[width=0.5\textwidth]{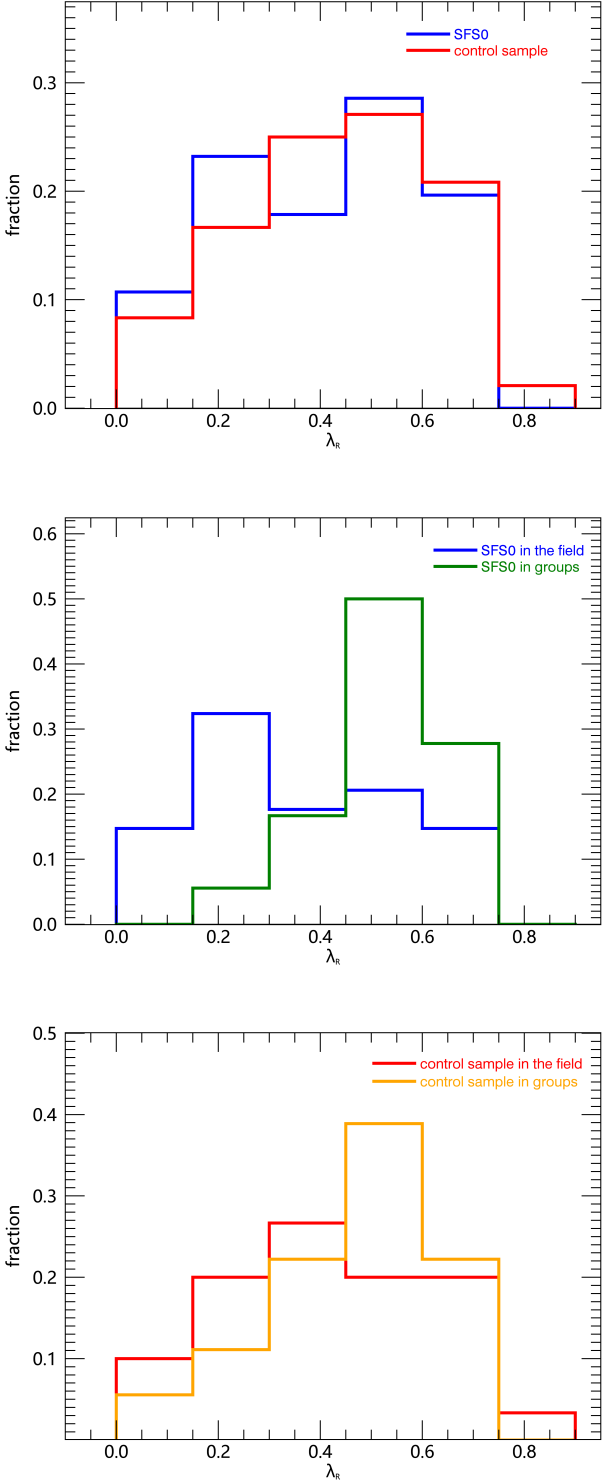}
\caption{The distributions of spin parameters. The top panel: SFS0 and the control sample. The middle panel: the SFS0s in the field and in groups. The bottom panel: the control sample in the field and in groups.\label{fig:7}}
\vspace{0.2cm}
\end{figure}

With MaNGA spatially resolved IFU data, we are able to calculate the spin parameter combining stellar kinematic measurements from individual spaxels. As discussed in Sec.~\ref{sec:dynamics}, we only use the spaxels within 45 degrees from major axis to avoid the deprojection effect near minor axis. We calculate all spaxels within 1.5\,R${\rm_e}$ and show the results in Fig.~\ref{fig:7}. The SFS0 and control sample have similar spin parameter distributions. To further investigate the effect of environments (e.g., the field and groups), we plot the distributions of $\rm\lambda_R$ for SFS0s and that of control sample in different environments separately in Fig.~\ref{fig:7}.\\
\indent Spin parameter distributions of SFS0s in the field and groups are different. In the field, they have average $\rm\lambda_R\sim0.40\pm0.04$, while in groups, the average of $\rm\lambda_R$ is $0.57\pm0.03$. However, there is no significant difference for the control sample. The control sample in groups is slightly more rotation-supported ($\rm\lambda_R\sim0.52\pm0.04$) than the isolated ($\rm\lambda_R\sim0.46\pm0.04$), which is consistent with \citet{2020MNRAS.498.2372D}. To confirm the validity of distinct distributions in SFS0s, we adopt the two-sample Kolmogorov-Smirnov test (K-S test). For SFS0s in the field and group, the p-value is $\sim$ 0.01 ($<$0.05), while for the control sample in the field and group, the p-value is $\sim$ 0.19. 
The above analysis implies that the SFS0s in the field and group may 
experience different dynamical processes, suggesting that the faded spiral scenario is more feasible in groups. Some isolated SFS0s exist having $\lambda_R > 0.45$, which will be discussed in Sec.~\ref{discussion}.

\subsection{N/O abundance}
\begin{figure}
\centering
\includegraphics[width=0.5\textwidth]{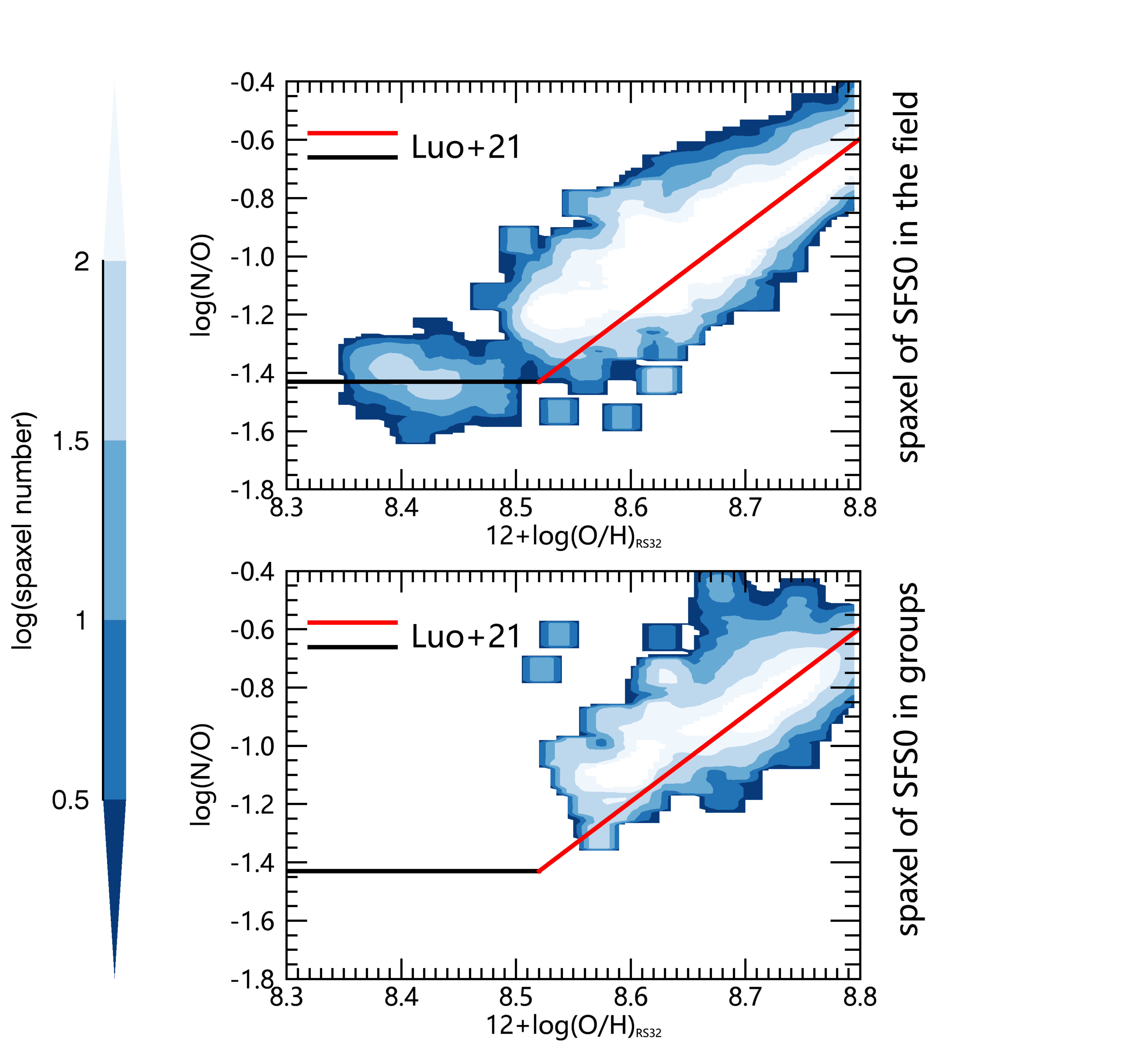}
\caption{The N/O vs O/H plane for SFS0s. The upper and lower panels are the spaxels from SFS0s in the field and in groups individually. The black and red solid lines are the result from \citet{2021ApJ...908..183L}. The color grid represents the spaxel number.\label{fig:10}}
\vspace{0.2cm}
\end{figure}

The method of N/O abundance ratio proposed by \citet{2021ApJ...908..183L} can provide evidence of accretion. We follow the same method to calculate the N/O and O/H for SFS0s (detailed in Sec.~\ref{sec:gasinflow}).\\
\indent In Fig.~\ref{fig:10}, the solid red line is the fitting for normal star-forming galaxies by \citet{2021ApJ...908..183L}. In this work, the abundance excess of N/O is not as high as \citet{2021ApJ...908..183L} because they selected the anomalously low-metallicity regions. But we still have several spaxels with 12+log(O/H) $<$ 8.50 around 
) $\sim $ -1.43. It can be seen that the excess of N/O is more obvious in the lower metallicity region (12+log(O/H) $<$ 8.65), especially for those spaxels in isolated SFS0s, the same region where the typical anomalously low-metallicity regions locate in \citet{2021ApJ...908..183L}. To better show the level of excess from the relation, we generate the Gaussian random points along the relation between log(N/O) and log(O/H). We assume that the random set has the same dispersion as SFS0 spaxels, $\sim$0.14. Then we apply the K-S test to the residuals of random Gaussian set and SFS0 spaxels from the relation. The p-value is smaller than $10^{-15}$ for both the spaxels of SFS0s in the field and groups. Therefore, the spaxels do not follow the normal galaxies, which possibly suggest that these regions have N/O abundance excess as those galaxies accreting metal-poor gas from the outside.
\vspace{0.6cm}

\section{Discussion}\label{discussion}
\subsection{Relation between $\rm M_*$ and metallicity}
\begin{figure}
\includegraphics[width=0.5\textwidth]{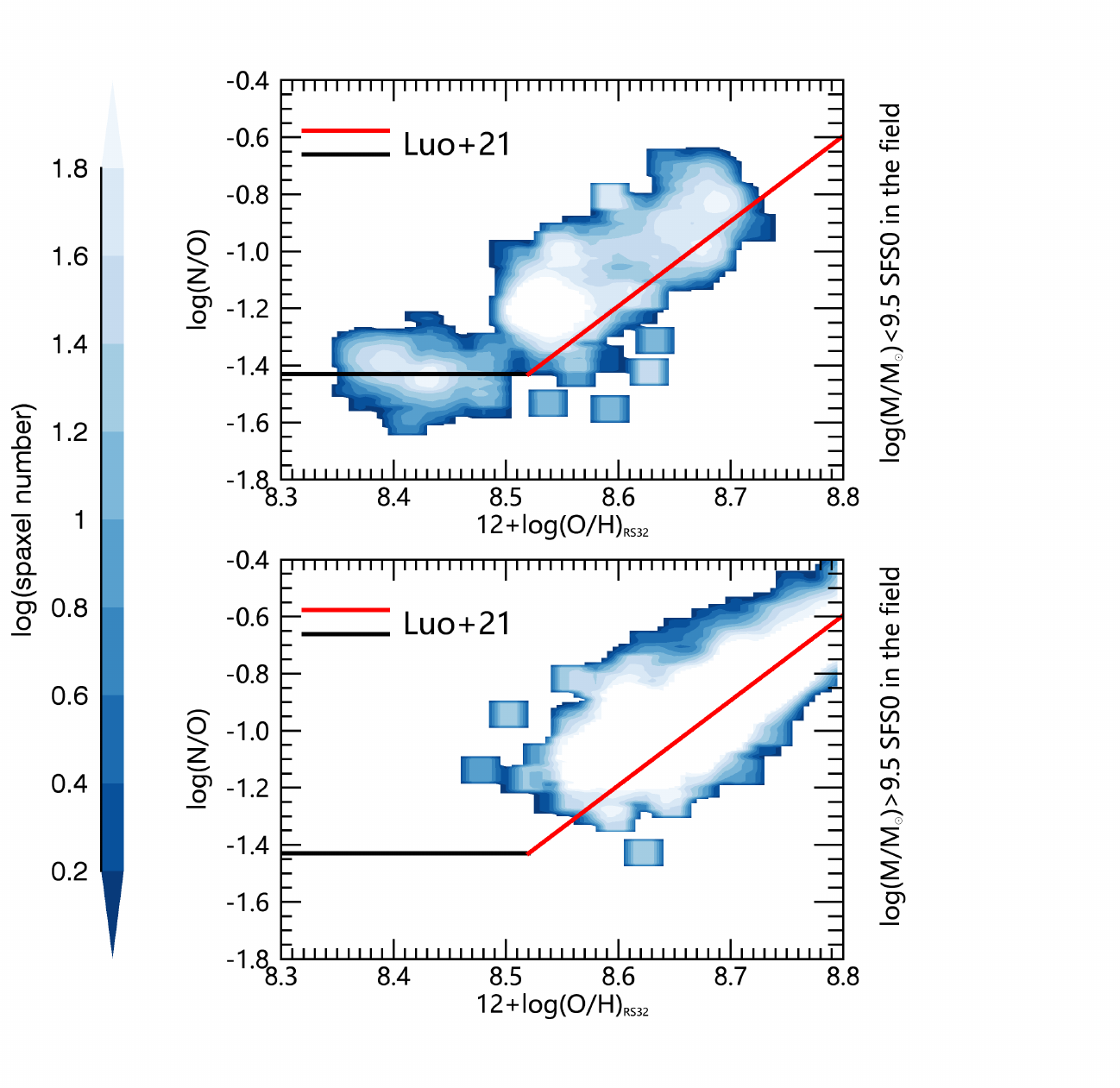}
\caption{The N/O vs O/H plane for the spaxels from low-mass (upper panel) and high-mass (lower panel) SFS0 in the field. The black and red solid lines are from \citet{2021ApJ...908..183L}. The color grid represents the spaxel number\label{fig:11}}
\vspace{0.2cm}
\end{figure}

\begin{figure}
\includegraphics[width=0.5\textwidth]{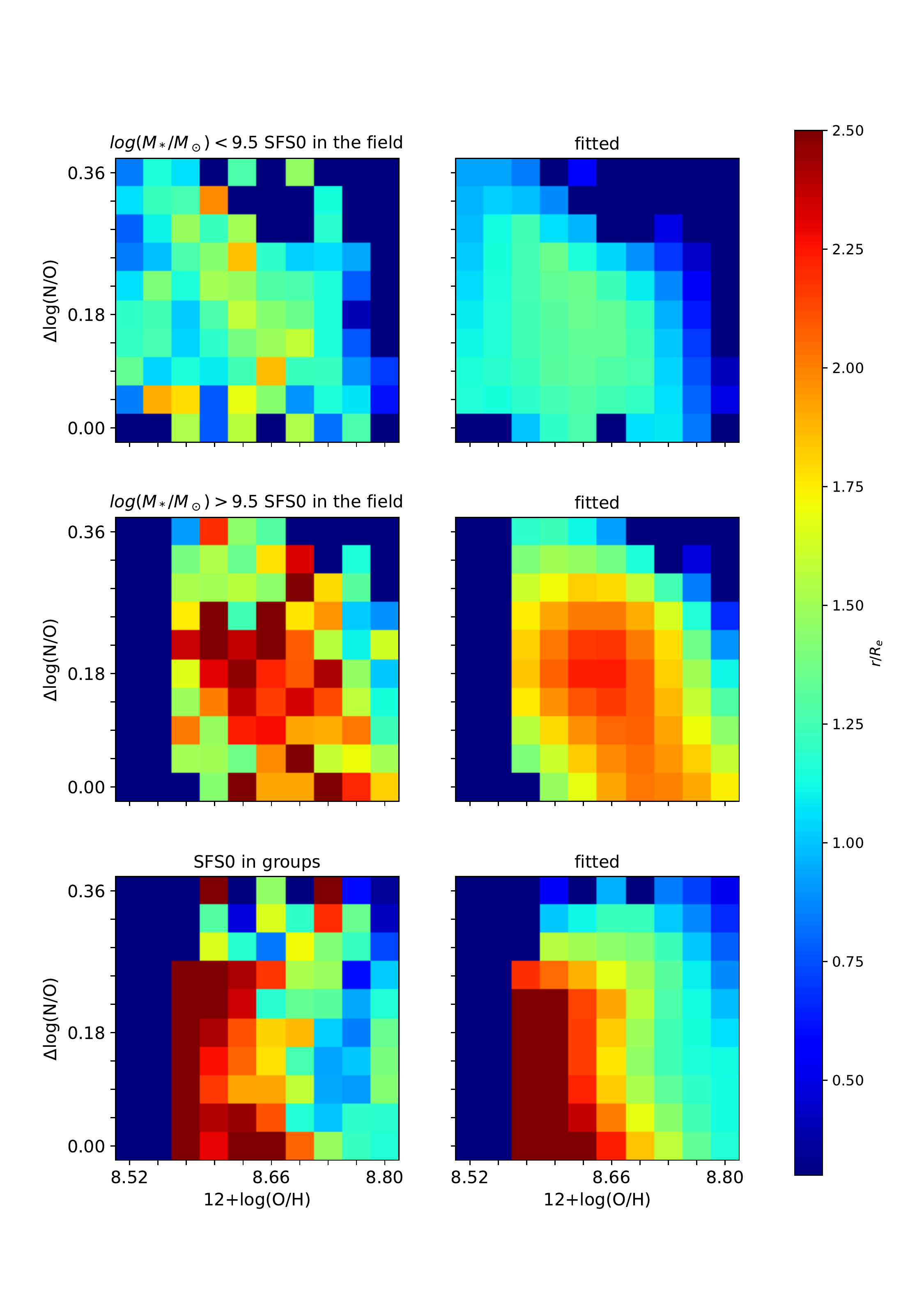}
\caption{The $\Delta$log(N/O) vs log(O/H) plane for the low-mass isolated SFS0s (top), high-mass isolated SFS0s (middle) and SFS0s in groups (bottom). $\rm\Delta log(N/O)=log(N/O)-log(N/O)_{Luo+21}$, where $\rm log(N/O)_{Luo+21}$ is calculated from the relation in \citet{2021ApJ...908..183L}. The left panels are the original calculation, and right panels are the results from `LOESS' fitting. The colour is the distance to the galactic centre normalized by $\rm R_e$.\label{fig:12}}
\vspace{0.2cm}
\end{figure}

Using a sample of $\sim$ 53,000 star-forming galaxies, \citet{2004ApJ...613..898T} found the relation between stellar mass and gas-phase metallicity, known as mass-metallicity relation (MZR). Recently, \citet{2016MNRAS.463.2513B} found that low-mass ($\rm\leq 10^{9.5}\;M_{\odot}$) galaxies deviated to lower metallicities by using MaNGA data. Therefore, it is necessary to examine the influence of stellar mass on metallicity for SFS0s.\\
\indent We divide SFS0s in the field into low-mass ones ($\rm log(M_*/M_{\odot})\leq9.5$; 9 galaxies) and high-mass ones ($\rm log(M_*/M_{\odot}>9.5$; 25 galaxies). The relation between N/O and O/H abundances are plotted in Fig.~\ref{fig:11}. The star-forming regions in low-mass isolated SFS0s have lower metallicity than the high-mass counterparts, aligned to the MZR. Besides, the low-mass ones locate in the expected regions in the N/O vs O/H plane, which indicates the possibility of metal-poor gas accretion. Since the merger history affects the specific angular momentum in the galaxy, a quick check is through their $\rm\lambda_R$, with 0.27$\pm$0.04 for the low-mass and 0.45$\pm$0.04 for the high-mass. The former shows disturbed kinematics, as expected after minor mergers \citep[e.g., ][]{2021MNRAS.502.3085G}. Even though the high-mass isolated SFS0s are more rotation-supported than the low-mass ones, they have smaller $\rm\lambda_R$ than SFS0s in groups, and we do not rule out the possibility of accretion or minor mergers. Because the galaxies absorbed by high-mass isolated SFS0s in minor mergers are also more massive and less metal-poor, no apparent nitrogen abundance excess is reasonable. Furthermore, they also make up the portions of isolated SFS0s with $\rm\lambda_R>0.45$ in Fig.~\ref{fig:7}.\\
\indent To find the spatial distribution of these spaxels, we set up the grid of $\rm\Delta log(N/O)$ vs log(O/H) ($\rm\Delta log(N/O)=log(N/O)-log(N/O)_{Luo+21}$, where $\rm log(N/O)_{Luo+21}$ is calculated from \citet{2021ApJ...908..183L}), and calculate the mean distance from the galaxy centre of the spaxels in the grid. The 12+log(O/H) range is restricted to [8.52,8.80] to contain enough spaxels. The calculated results are then fitted with the `LOESS' procedure \citep{2013MNRAS.432.1862C} and shown in Fig.~\ref{fig:12}. The `LOESS' procedure is used here to recover the trends from the noisy data.\\
\indent The trend of SFS0s in groups can be explained by a negative radial metallicity gradient where low-metallicity spaxels locate in a distant region. For high-mass isolated SFS0s, most spaxels have a mean distance of 2 $\rm R_e$, indicating the existence of the young stellar population at larger radii. 
\citet{2021ApJ...908..183L} found the highest N/O excess region at $\rm r>1.5\;R_e$. If we suppose that these high-mass isolated SFS0s accrete materials from outside, our results are similar to \citet{2021ApJ...908..183L}
and the relatively high angular momentum help stabilize the disc and prevent gas inflowing \citep[e.g. ][]{2016ApJ...824L..26O}. For the low-mass isolated SFS0s, we find the spaxels with nitrogen abundance excess in the central region, which provides evidence for the metal-poor gas inflow fueling the central star formation. We note that it is an average result not standing for a single galaxy.


\subsection{SF quenching or rejuvenating}\label{qorr}
The environment may have important imprints on the galaxy evolution in addition to the morphological quenching \citep[e.g.][]{2020ApJ...902...75L}. For example, the anomalously low-metallicity galaxies in \citet{2021ApJ...908..183L} accrete metal-poor gas from the circumgalactic medium or the cosmic web. For galaxies in groups or the field, the ongoing physical process can be very different. In the field, the minor mergers and inflow of fresh gas are common for S0s \citep{2014MNRAS.440.3491J}, while in the denser environment, S0s are often found in the outskirt of clusters, when the harassment \citep[e.g.][]{1996Natur.379..613M}, tidal interactions \citep[e.g.][]{2006MNRAS.369.1021M} and stripping \citep[e.g.][]{2020MNRAS.495..554R,1982MNRAS.198.1007N} are important, especially as falling towards the central region of clusters.  \\
\indent In the local universe, minor mergers happen frequently \citep[e.g.][]{1988ApJ...327..507F,2004MNRAS.355..819G}. \citet{2014MNRAS.437L..41K,2014MNRAS.440.2944K} suggested that almost half of nearby star formation activities are triggered by minor mergers. Because minor mergers are not destructive, unlike major mergers, the disc structure can be preserved. Minor mergers triggering star formation has been observed in S0s \citep[e.g.][]{2020ApJ...889..132G,2021ApJ...915....1C}. On the other hand, in numerical simulations, minor mergers are shown to thicken the disc and decrease the specific angular momentum \citep[e.g.][]{2021MNRAS.502.3085G}. After the minor merger, the angular momentum should be reduced from the initial state, explaining the low $\lambda_R$ distribution in isolated SFS0s. For those high-mass isolated SFS0s with higher $\lambda_R$, as shown in Fig.~\ref{fig:11} that the accretion is not remarkable, they can reserve the relatively high angular momentum. The thickness and disturbances of the disc would be a good tracer of mergers, which is of further interest with deep imaging.\\
\indent On the other hand, in groups, processes like ram pressure stripping could make the infalling satellites more gas-poor in denser environments \citep{2011ApJ...729...11K}. The environmental effects would affect the material cycle in the galaxy \citep{2017ARA&A..55..389T} and strip the atomic gas, which is less bounded to the host. However, during the process, such as stripping, the star formation rate can also be enhanced \citep[e.g.][]{2020MNRAS.495..554R}. For normal spiral galaxies, the gas depletion time is $\sim$ 1.0 Gyr considering the Kennicutt-Schmidt law, which is much shorter than the cosmic age. Within dense environments, the depletion would be more effective. From the statistics, we would observe fewer such SFS0s in groups than in the field, in coincidence with our work (18 in groups and 34 in the field). In contrast, normal S0 galaxies have similar numbers in different environments. Because the faded spirals are mainly the predecessors of S0s in groups or clusters \citep{1951ApJ...113..413S,1972ApJ...176....1G,1980ApJ...236..351D,1997ApJ...490..577D}, they are more rotation-supported. They have low-metallicity regions at larger radii, implying a stable disc resisting gas inflow. 
With more data and larger samples in the future, we can further study the impacts of group richness and distances from the group centre on SFS0s.\\
\indent Compared with \citet{2021ApJ...908..183L}, our SFS0 sample has larger O/H in their spaxels, which is certain because their sample consist of low-metallicity regions. With the old evolved stellar population in early-type galaxies, it is not strange that the nitrogen abundance excess is not obvious. Another possibility is that the change in metallicity might depend on the properties of the interacting neighbourhood \citep[e.g. ][]{2021arXiv210805014P}. We note that, due to the old stellar population in these early-type galaxies, we will not have the problem of Wolf-Rayet stars as \citet{2021ApJ...908..183L}. Our sample selection excludes the AGNs and has moderate star formation activities, which makes the outflow impossible to dilute the ISM.\\
\indent \citet{2013ApJ...765..140A} found N/O $\propto$ O/H$^{1.7}$ using the global data. Our SFS0s in groups are consistent with their slope in Fig.~\ref{fig:10}, despite the different intercepts. The construction of global N/O abundances from the resolved ones and the detailed chemical evolution for SFS0s are of future interest.

\subsection{Spatially resolved SFMS and star formation law}\label{sec:resolved}
As discussed in Sec.~\ref{sec:rSFMS}, rSFMS with the slope $\sim$ 1.0 obtained in this work is consistent with the previous work using different data sets. The slope of $\sim$ 1.0 enables the construction of global SFMS from the resolved scales, though the scatter of rSFMS could contribute to the non-linear global relation. It also implies that SFS0s regulate their star formation activities at resolved scales of kpc.\\
\indent To derive the SFMS, given the Kennicutt-Schmidt law, we can take into consideration the molecular gas main sequence \citep[MGMS, e.g.,][]{2021MNRAS.503.1615S}, the molecular gas fraction, the free-fall time, and a dimensionless parameter $\epsilon_{ff}$, which describes the star formation efficiency per free-fall time \citep[e.g.][]{2005ApJ...630..250K,2012ApJ...745...69K,2018MNRAS.477.2716K}. Because star formation relates tightly to the dense gas \citep{2004ApJ...606..271G}, the cold gas fraction could be responsible for the scatter in the SFMS. On the other hand, there are different modified Kennicutt-Schmidt laws. E.g., \citet{2011ApJ...733...87S,2018ApJ...853..149S} proposed the modified Kennicutt-Schmidt law as:
\begin{equation}
\Sigma_{\rm SFR}\propto (\Sigma_{\rm star}^{0.5}\Sigma_{\rm gas})^{1.09}.
\end{equation}
However, the volumetric star formation law proposed by \citet{2012ApJ...745...69K} is opposed to \citet{2011ApJ...733...87S,2018ApJ...853..149S} in some ETG sample \citep[which has low star formation efficiency but large stellar mass surface density, e.g.,][]{2014MNRAS.444.3427D} or some dwarf galaxies \citep[which have low stellar mass surface density, e.g.,][]{2018ApJ...853..149S}. Thus, it is hard to construct the SFMS through a simple star formation law. Unlike spirals, S0s do not have spiral patterns, and the bulge would play a more prominent role in shearing and preventing the gas from infalling \citep{2014MNRAS.444.3427D}. But as in Sec.~\ref{sec:cs}, most of our SFS0s have pseudo bulges, indicating that the bulges are not able to prevent gas infall \citep{1994ApJ...437L..47M,1995ApJ...448...41H}. Additionally, the specific angular momentum also influences the atomic content in a disc galaxy \citep[e.g.,][]{2019MNRAS.483.2398M,2017MNRAS.467.1083L}, which further influences the reservoir of cold gas. For a better knowledge of S0s, cold gas data are essential. Unfortunately, we fail to match a sufficient number of galaxies in the HI blind survey ALFALFA \citep{2018ApJ...861...49H}. We have proposed the observation with JCMT for SFS0s. If we obtained the gas data, we could better understand the star formation laws for S0s.

\section{Conclusion}\label{sec:conclusion}
In this work, we select the S0 sample from the previous deep learning results for the SDSS-IV MaNGA survey. Our main results are the following:

(1) The SFS0s show a smaller mean bulge S\'{e}rsic index.

(2) The isolated SFS0s have smaller $\lambda_R$ than SFS0s in groups. Compared to SFS0s, the normal S0s do not show obvious differences in different environments.

(3) The resolved star formation main sequence has a slope $\sim$ 1.0, consistent with normal star-forming galaxies.

(4) The isolated SFS0 spaxels show more obvious N/O excess than SFS0s in groups, which might be due to accretions or minor mergers.

(5) The low-mass isolated SFS0s have similar N/O abundances as anomalously low-metallicity regions in \citet{2021ApJ...908..183L} compared to the high-mass ones.

\indent From above, we provide clues for the origin of SFS0s. In groups, they are possible successors of faded spirals. In the field, minor mergers could provide metal-poor gas and ignite star formation in S0s, especially in low-mass ones. And their star formation is self-regulated at kpc scales during quenching or rejuvenation, identical to normal star-forming galaxies.

In future research, we expect observation results from JCMT and to find more information with gas data. Besides, with the coming MaNGA new data release, we would have a larger sample and high quality data.

\section*{Acknowledgements}
The authors are very grateful to the anonymous referee for critical comments and instructive suggestions, which significantly strengthened the analysis in this work.\\
\indent This work is supported by the National Key Research and Development Program of China (No. 2017YFA0402703) and by the National Natural Science Foundation of China (No. 11733002).\\
\indent Funding for the Sloan Digital Sky Survey IV has been provided by the Alfred P. Sloan Foundation, the U.S. Department
of Energy Office of Science, and the Participating Institutions. SDSS-IV acknowledges support and resources from the centre for High-Performance Computing at the University of Utah. The SDSS web site is www.sdss.org. SDSS-IV is managed by the Astrophysical Research Consortium for the Participating Institutions of the SDSS Collaboration including the Brazilian Participation Group, the Carnegie Institution for Science, Carnegie Mellon University, the Chilean Participation Group, the French Participation Group, Harvard-Smithsonian centre for Astrophysics, Instituto de Astrof\'{i}sica de Canarias, The Johns Hopkins University, Kavli Institute for the Physics and Mathematics of the Universe (IPMU) / University of Tokyo, Lawrence Berkeley National Laboratory, Leibniz Institut f$\rm \ddot{u}$r Astrophysik Potsdam (AIP), Max-Planck-Institut f$\rm \ddot{u}$r Astronomie (MPIA Heidelberg), Max-Planck-Institut f$\rm \ddot{u}$r Astrophysik (MPA Garching), Max-Planck-Institut f$\rm \ddot{u}$r Extraterrestrische Physik (MPE), National Astronomical Observatory of China, New Mexico State University, New York University, University of Notre Dame, Observat\'{o}rio Nacional / MCTI, The Ohio State University, Pennsylvania State University, Shanghai Astronomical Observatory, United Kingdom Participation Group, Universidad Nacional Aut\'{o}noma de M\'{e}xico, University of Arizona, University of Colorado Boulder, University of Oxford, University of Portsmouth, University of Utah, University of Virginia, University of Washington, University of Wisconsin, Vanderbilt University, and Yale University.\\

\section*{Data Availability}
The DAP and PIPE3D pipeline data are available via https://www.sdss.org/dr16/manga/. And the procedure ‘LOESS’ is available via http://wwwastro.physics.ox.ac.uk/ mxc/software/ and \citet{2013MNRAS.432.1862C}.



\bibliographystyle{mnras}





\bsp	
\label{lastpage}
\end{document}